\documentclass[12pt]{article}
\usepackage[a4paper,tmargin=3truecm,bmargin=3truecm,rmargin=2.2truecm,lmargin=2.2truecm]{geometry}
\usepackage{epsf}
\usepackage{amssymb}
\usepackage{amsmath}
\usepackage{amsfonts}
\usepackage{cite}
\usepackage[small]{caption2}

\usepackage[T1]{fontenc}
\usepackage{psfrag,epsfig,graphicx,graphics}

\newcommand{\scalings}{1.4cm}
\newcommand{\scalingsb}{1.62cm}
\newcommand{\scalingm}{2.2cm}

\newcommand{\scalingml}{2.6cm}
\newcommand{\scaling}{2.98cm}
\newcommand{\scalingb}{3.2cm}
\newcommand{\scalingl}{4cm}

\newcommand{\scalingA}{2.3cm}
\newcommand{\scalingB}{2.8cm}
\newcommand{\scalingC}{3.2cm}
\newcommand{\scalingD}{2.9cm}

\newcommand{\be}{\begin{equation}}
\newcommand{\beq}{\begin{equation}}
\newcommand{\ee}{\end{equation}}

\newcommand{\bea}{\begin{eqnarray}}
\newcommand{\eea}{\end{eqnarray}}

\newcommand{\lu}{$\ell_1$}
\newcommand{\ld}{$\ell_2$}
\newcommand{\lut}{$\tilde{\ell}_1$}
\newcommand{\ldt}{$\tilde{\ell}_2$}

\def\slashchar#1{\setbox0=\hbox{$#1$}
   \dimen0=\wd0
   \setbox1=\hbox{/} \dimen1=\wd1
   \ifdim\dimen0>\dimen1
      \rlap{\hbox to \dimen0{\hfil/\hfil}}
      #1
   \else
      \rlap{\hbox to \dimen1{\hfil$#1$\hfil}}
      /gdatdafinal2.tex
   \fi}

\begin{document}

\begin{flushright}
CPHT-RR-024.0506 \\
LPT 06-15 \\
\end{flushright}

\vspace{\baselineskip}

\begin{center}

\textbf{\LARGE QCD factorizations in
$\gamma^* \gamma^* \to \rho^0_L \rho^0_L$}\\
\vspace{3\baselineskip}


{\large
B.\ Pire$^a$,
M.\ Segond$^{b}$,
L.\ Szymanowski$^{b,c,d}$ and
S.\ Wallon$^b$
}\\

\vspace{2\baselineskip}

${}^a$\,CPHT\footnote{Unit{\'e} mixte 7644 du CNRS},
\'Ecole Polytechnique, 91128 Palaiseau, France \\[0.5\baselineskip]
${}^b$\,LPT\footnote{Unit{\'e} mixte 8627 du CNRS}, Universit\'e
Paris-Sud, 91405-Orsay, France \\[0.5\baselineskip]
${}^c$\,Universit\'e  de Li\`ege,  B4000  Li\`ege,
Belgium\\[0.5\baselineskip]
${}^d$\,So{\l}tan Institute for Nuclear Studies, Ho\.za 69, 00-681 Warsaw, Poland\\[0.5\baselineskip]

\vspace{5\baselineskip}
\textbf{Abstract}\\
\vspace{1\baselineskip}
\parbox{0.9\textwidth}{We calculate the lowest order QCD amplitude, i.e. the quark exchange
contribution,
 to the forward production amplitude of a pair of longitudinally polarized
$\rho$ mesons in the scattering of two virtual photons
$\gamma^*(Q_1) \gamma^*(Q_2) \to \rho^0_L \rho^0_L$. We show that
the scattering amplitude simultaneously factorizes in two quite
different ways:
 the part with  transverse photons is described by the QCD factorization formula
involving the generalized distribution amplitude of two final $\rho$
mesons, whereas the part with longitudinally polarized photons takes
the QCD factorized form with the
 $\gamma^*_L \to \rho^0_L$ transition distribution amplitude.
Perturbative expressions for these, in general,
 non-perturbative functions are obtained in terms of the $\rho-$meson distribution amplitude.
}

\end{center}

\setcounter{footnote}{0}

\newpage

\section{Introduction}
\begin{figure}[htb]
\psfrag{r1}[cc][cc]{$\quad\rho(k_1)$}
\psfrag{r2}[cc][cc]{$\quad\rho(k_2)$}
\psfrag{p1}[cc][cc]{$\slashchar{p}_1$}
\psfrag{p2}[cc][cc]{$\slashchar{p}_2$}
\psfrag{q1}[cc][cc]{$q_1$}
\psfrag{q2}[cc][cc]{$q_2$}
\psfrag{Da}[cc][cc]{DA}
\psfrag{HDA}[cc][cc]{$M_H$}
\psfrag{M}[cc][cc]{$M$}
\centerline{\scalebox{1}
{
$\begin{array}{cccc}
\raisebox{-0.44 \totalheight}{\epsfig{file=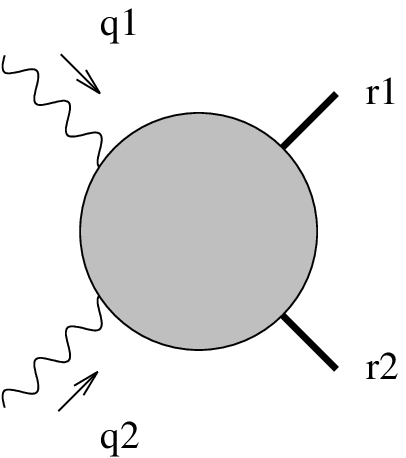,width=\scalingl}}&=&
\raisebox{-0.44\totalheight}{\epsfig{file=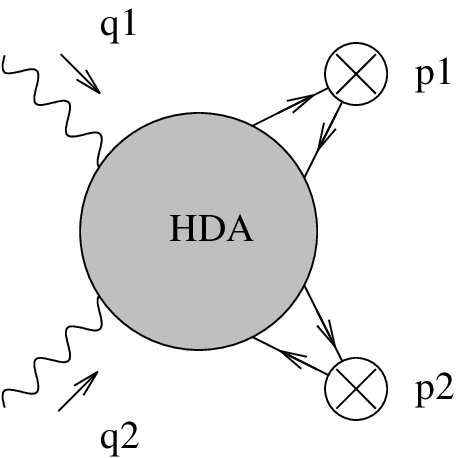,width=\scalingl}}
& \begin{array}{c}
\raisebox{0.4 \totalheight}
{\psfrag{r}[cc][cc]{$\quad\rho(k_1)$}
\psfrag{pf}[cc][cc]{$\slashchar{p}_2$}
\epsfig{file=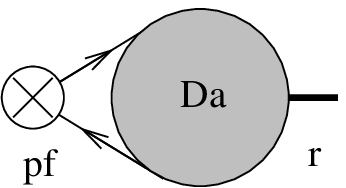,width=\scalingl}}\\
\raisebox{0.1 \totalheight}
{\psfrag{r}[cc][cc]{$\quad\rho(k_2)$}
\psfrag{pf}[cc][cc]{$\slashchar{p}_1$}
\epsfig{file=DA.eps,width=\scalingl}}
\end{array}
\end{array}
$}}
\caption{\small The amplitude of the process $\gamma^*(Q_1) \gamma^*(Q_2) \to \rho^0_L (k_{1})\rho^0_L(k_{2})$ in
the collinear factorization. \label{M}}
\end{figure}

The exclusive reaction \beq \label{proces} \gamma^*(q_1)
\gamma^*(q_2) \to \rho^0_L (k_{1})\rho^0_L(k_{2}). \ee is a
beautiful theoretical laboratory for the understanding of
factorization properties of high energy QCD. In  previous  studies
we emphasized
 the gluon exchange contribution which dominates in the very high energy (small-x) regime, both at the
 Born order \cite{PSW} and in the resummed BFKL approach \cite{EPSW}. At lower scattering energy
one can expect a non negligible contribution of quark exchange,
since  in the strong coupling $g$ expansion, quark exchange
processes appear at lower order, namely at order $g^2$,  than the
gluon exchange ones.
 The study of these quark exchanges processes  is the main
motivation of the present work. The Born order contribution with quark exchanges
is described by the
 same set of diagrams  which contribute to the  scattering of real photons
producing pions, e.g. $\gamma \gamma \to \pi^+\pi^-$, at large
momentum transfer, studied long ago by Brodsky and Lepage
\cite{BLphysrev24} in the framework of the factorized form of
exclusive processes at fixed angle \cite{ERBL}, in which mesons are
described by their light-cone distribution amplitudes (DAs). This
is illustrated in Fig.\ref{M}.
In this sense our present study can be seen as a complement of Ref.
\cite{BLphysrev24} for the case of the scattering with virtual
photons, i.e. with both, transverse and longitudinal polarizations,
and in the
 forward kinematics (see also Ref. \cite{DFKV}).
The virtualities $Q^2_i=-q^2_i$, $i=1,2$, supply the hard scale to
the process (\ref{proces}) which justifies the use of the QCD
collinear  factorization methods and the description of $\rho$
mesons by means of their distribution amplitudes.

We calculate in Sec.~\ref{secborn} within this scheme the scattering
amplitude of (\ref{proces}) at Born level. At this stage we do not
impose any additional conditions on the magnitudes of photon
virtualities $Q_i$. Next,
 we  turn to the study
 of two particular kinematical regions:  (1) the region where the
 squared invariant mass of the two rho's $W^2$ is much smaller than the largest photon virtualities,
 namely $Q_1^2 \gg W^2$ (or $Q_2^2 \gg W^2$), with $Q_1$ and $Q_2$ being not parametrically close, 
see Sec.~\ref{secgda} for more details,
 and (2) the region where
photon virtualities are strongly ordered, that is $Q_1^2 >> Q_2^2$ (or $Q_2^2
>> Q_1^2)$. We will show in Sec.~\ref{secgda} that in the region (1)
the amplitude with transverse photons factorizes in a hard subprocess and a Generalized
Distribution Amplitude \cite{DM,GDA}, up to corrections of order
$W^2/Q_1^2$ (resp. $W^2/Q_2^2$). In the region (2) the amplitude
with longitudinal photons
factorizes in a hard subprocess and a Transition Distribution
Amplitude \cite{TDA}, up to corrections of order $Q_2^2/Q_1^2$
(resp. $Q_1^2/Q_2^2$), as is shown in Sec.~\ref{sectda}.
\begin{figure}[htb]
\begin{picture}(500,340)
\put(0,0){\epsfxsize=12cm{\centerline{\epsfbox{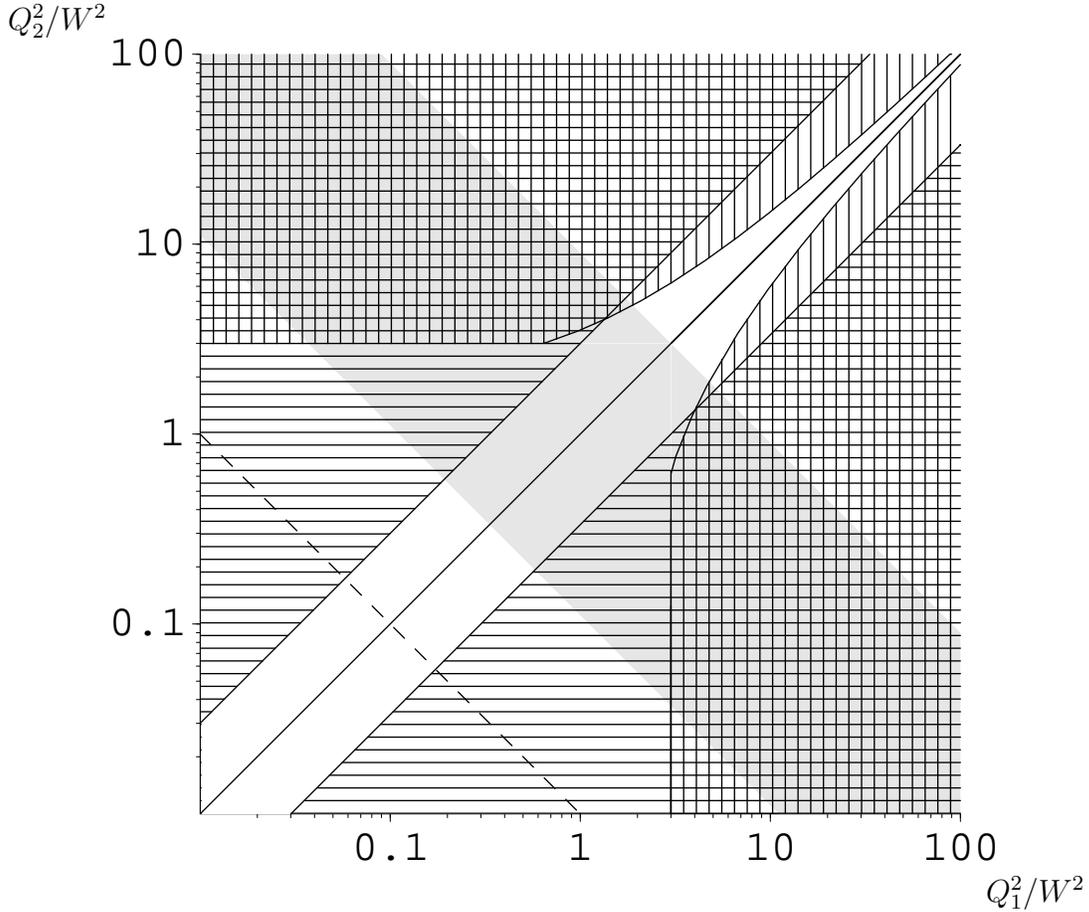}}}}
\put(400,0){$Q_1^2/W^2$}
\put(30,330){$Q_2^2/W^2$}
\end{picture}
\caption{Different kinematical regions.  
In the domain (1), denoted with vertical  lines, the QCD factorization with a
GDA is justified. This region is the union of two disconnected domains: the
lower right one is given by the conditions $1-Q_2^2/s \ge c(1-Q_1^2/s)$ and
$Q_1^2/W^2 \ge c$ while the upper left one is given by $1-Q_1^2/s \ge
c(1-Q_2^2/s)$ and $Q_2^2/W^2 \ge c,$ $c$ being arbitrary large.
In the domain (2), denoted with horizontal lines, the QCD factorization with a TDA is justified. It corresponds to $Q_1^2/Q_2^2 \ge c$ or $Q_1^2/Q_2^2  \le 1/c.$
In the intersecting domain, both factorizations are valid, while in the  region without any lines, no
 factorization neither in terms of GDA nor TDA is established. 
For illustration, we choose $c=3.$ The grey band represents the domain
  where the Born order amplitudes with transverse and  longitudinal photon,  calculated directly as in Sec.~\ref{secborn}, have comparable magnitudes.
 In the upper (lower) corner the transverse (resp. longitudinal) polarizations give dominant contribution. Below the
 dashed line is the perturbative Regge domain.
\label{Figregions}}
\end{figure}
These two domains have a non empty intersection as shown in
Fig.\ref{Figregions}.
 In this intersection, we get two factorisation formulas
 for both polarizations of the photons,
 $\gamma^*_T \gamma^*_T \to \rho_L \rho_L$ and  $\gamma^*_L \gamma^*_L \to \rho_L \rho_L.$ 
Fig.~\ref{Figregions} illustrates also that the collinear QCD
factorization with GDA or TDA is not demonstrated by our analysis in the limited
region where the virtualities $Q_i^2$ are parametrically close. In this figure, 
 the different domains where the transverse or longitudinal
amplitudes dominates are displayed.
Finally, Fig.~\ref{Figregions} shows the perturbative Regge domain,
corresponding to  the large $W^2$ limit.

 In this paper we concentrate ourselves, for simplicity,
on the case of longitudinally polarized rho mesons. In this spirit,
the longitudinal rho pairs and the pion pairs lead to similar
results.
 Our study is closely related to the perturbative limit of
the two-pion light-cone distribution amplitude performed in
Ref.\cite{DFKV} for $\gamma^*_T \gamma_T$ collisions in the GDA
limit.

Thus, the question we investigate here is how the known QCD
factorization formulas emerge from a Born order
calculation in both the longitudinally polarized and transversally polarized photon cases.
As a by product, we get a perturbative expression for the non-perturbative hadronic objects - the $\rho\,\rho$ GDA and
the $\gamma^* \to \rho$ TDA - in terms of the $\rho-$meson distribution amplitude.

\section{Kinematics}
Let us fix the kinematics, which is quite simple in the forward regime that we are investigating. The inclusion of transverse momentum would not change
dramatically  our result but would necessitate the introduction of more complicated tensorial structures.
We use a Sudakov
decomposition with two lightlike vectors $p_1$ and $p_2$ with $2 p_1 . p_2 = s$ and write the photon momenta as
\beq
q_{1} = p_1 - \frac{Q_{1}^2}{s} p_2 ~~~~~~~~q_{2} = p_2 - \frac{Q_{2}^2}{s} p_1\;,
\ee
and the final meson momenta as
\beq
k_{1} = (1  - \frac{Q_{2}^2}{s}) p_1 ~~~~~~~~k_{2} = (1  - \frac{Q_{1}^2}{s}) p_2\;.
\ee
The positivity of energy of produced $\rho'$s requires that $s \geq Q_i^2$.
The usual invariant $W^2$ is defined as
\beq
\label{Ws}
W^2 = (q_{1}+ q_{2})^2 = (k_{1}+ k_{2})^2 =s  (1  - \frac{Q_{1}^2}{s}) (1  - \frac{Q_{2}^2}{s})\;,
\ee
or
\beq
\label{sW}
s=\frac{1}{2}\left[Q_1^2 +Q_2^2 +W^2 + \lambda(Q_1^2,Q^2_2,-W^2)   \right]\;,
\ee
with
$\lambda(x,y,z)\equiv \sqrt{x^2+y^2+z^2-2xy-2xz-2zy}\;$,
while the minimum squared momentum transfer is
\beq
t_{min} = (q_{2}- k_{2})^2 = (q_{1}+ k_{1})^2 = - \frac{Q_{1}^2 Q_{2}^2}{s}\;.
\ee
Note that, contrarily to the case studied in Ref.\cite{BLphysrev24,DFKV },  $t_{min}$ may not be large with respect to $\Lambda^2_{QCD}$. depending on the respective values of
$Q_1^2$, $Q_2^2$ and $W^2$.

\section{The Born order amplitude}
\label{secborn}

The scattering amplitude ${\cal A}$ of the process (\ref{proces}) can be written in the form
\beq
\label{scatampl}
{\cal A}= T^{\mu\, \nu}\epsilon_\mu(q_1)\epsilon_\nu(q_2)\;,
\ee
where the tensor $T^{\mu\, \nu}$ has in the above kinematics a simple decomposition which is consistent with
  Lorentz covariance and  electromagnetic gauge invariance
\beq
\label{T}
T^{\mu\, \nu} = \frac{1}{2}g^{\mu\,\nu}_T\;(T^{\alpha\, \beta}g_{T\,\alpha\,\beta}) +
(p_1^\mu +\frac{Q_1^2}{s}p_2^\mu)(p_2^\nu
+ \frac{Q_2^2}{s}p_1^\nu)\;\frac{4}{s^2}(T^{\alpha\, \beta}p_{2\,\alpha}\, p_{1\,\beta})\;,
\ee
and where $g^{\mu\,\nu}_T=g^{\mu \nu} - (p_1^\mu p_2^\nu + p_1^\nu p_2^\mu)/(p_1 . p_2)$.
The first term on the rhs of Eq.~(\ref{T}) contributes in the case of transversely polarized photons,
the second one for longitudinally polarized virtual photons.
The longitudinally polarized
$\rho^0-$meson DA $\phi(z)$ is defined by the non-local correlator (for simplicity of notation we omit
the Wilson line)
\be
\label{DArho0}
\langle \rho^0_L(k)|\bar q(x)\gamma^\mu q(0)|0\rangle =
\frac{f_\rho}{\sqrt{2}}\; k^\mu \;\int\limits_0^1\,dz\,e^{iz(kx)}\phi(z)\;, \;\;\;\;\;\mbox{for}\;\;q=u,d\;,
\ee
proportional to the coupling constant $f_\rho$.

\begin{figure}[htb]
\psfrag{l1}[cc][cc]{\lu}
\psfrag{l2}[cc][cc]{\ld}
\psfrag{l1t}[cc][cc]{\lut}
\psfrag{l2t}[cc][cc]{-\ldt}
\psfrag{q1}[cc][cc]{$q_1$}
\psfrag{n}[cc][cc]{}
\begin{eqnarray*}
\begin{array}{ccccccc}
\!\!\!\!\left(\raisebox{-0.44 \totalheight}{\epsfig{file=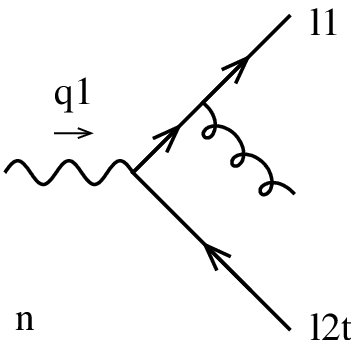,width=\scaling}}\right.&+&\left.
\raisebox{-0.44 \totalheight}{\epsfig{file=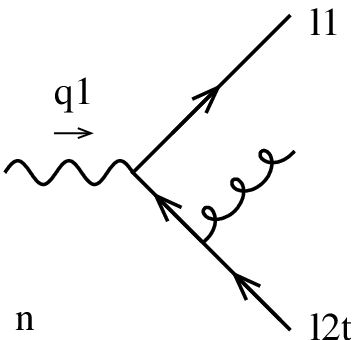,width=\scaling}}\,\right)
&\otimes&
\left(\raisebox{-0.44 \totalheight}
{\psfrag{q1}[cc][cc]{$q_2$}
\psfrag{l1}[cc][cc]{-\lut}
\psfrag{l2t}[cc][cc]{\ld}
\epsfig{file=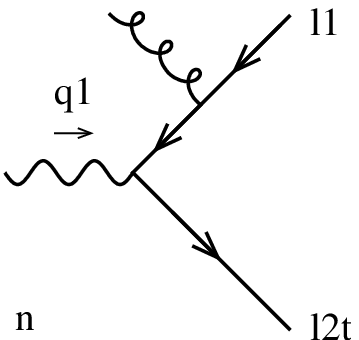,width=\scaling}}\right.&+&\left.\raisebox{-0.44 \totalheight}
{\psfrag{q1}[cc][cc]{$q_2$}
\psfrag{l1}[cc][cc]{-\lut}
\psfrag{l2t}[cc][cc]{\ld}
\epsfig{file=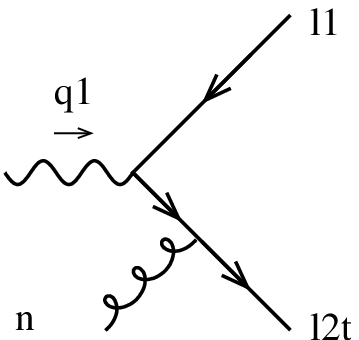,width=\scaling}}\,\right) \\
&&&&&& \\
\!\!\!\!(1a) & & (1b) & & (2a) & & (2b) \\
\!\!\!\!&&&+&&& \\
\!\!\!\!&&&&&& \\
\!\!\!\!\left(\raisebox{-0.44 \totalheight}
{\psfrag{q1}[cc][cc]{$q_2$}
\epsfig{file=diag1.eps,width=\scaling}}\right.&+&\left.\raisebox{-0.44 \totalheight}
{\psfrag{q1}[cc][cc]{$q_2$}
\epsfig{file=diag2.eps,width=\scaling}}\,\right)
&\otimes& \left(\raisebox{-0.44 \totalheight}
{\psfrag{l1}[cc][cc]{-\lut}
\psfrag{l2t}[cc][cc]{\ld}
\epsfig{file=diag3.eps,width=\scaling}}\right.&+&\left.\raisebox{-0.44 \totalheight}
{\psfrag{l1}[cc][cc]{-\lut}
\psfrag{l2t}[cc][cc]{\ld}
\epsfig{file=diag4.eps,width=\scaling}}\,\right) \\
\!\!\!\!&&&&&& \\
\!\!\!\!(3a) & & (3b) & & (4a) & & (4b) \end{array}\\
\end{eqnarray*}
\caption{Feynman diagrams contributing to $M_H$,
in which the virtual photons couple to  different quark lines. \label{diffDiagrams}}
\end{figure}

\begin{figure}[htp]
\begin{eqnarray*}
\psfrag{q1}[cc][cc]{$q_1$}
\psfrag{q2}[cc][cc]{$q_2$}
\psfrag{l1}[cc][cc]{\lu}
\psfrag{l1t}[cc][cc]{-\lut}
\psfrag{l2}[cc][cc]{\ld}
\psfrag{l2t}[cc][cc]{\!-\ldt}
\psfrag{n}[cc][cc]{}
\begin{array}{ccccccccccccc}
\left(\raisebox{-0.44 \totalheight}{\epsfig{file=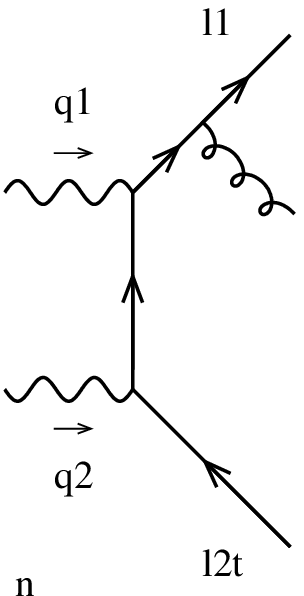,width=\scalings}}\right.&\!+&\!\raisebox{-0.44 \totalheight}{\epsfig{file=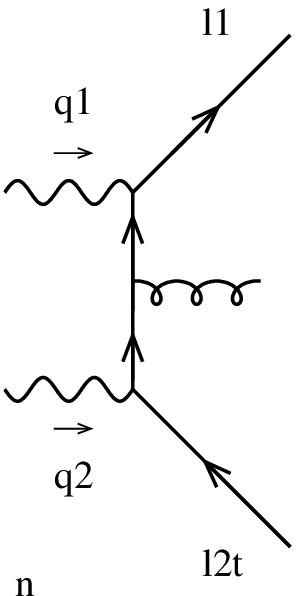,width=\scalings}}&\!+&\!\raisebox{-0.44 \totalheight}{\psfrag{l2t}[cc][cc]{\!\!-\ldt}\epsfig{file=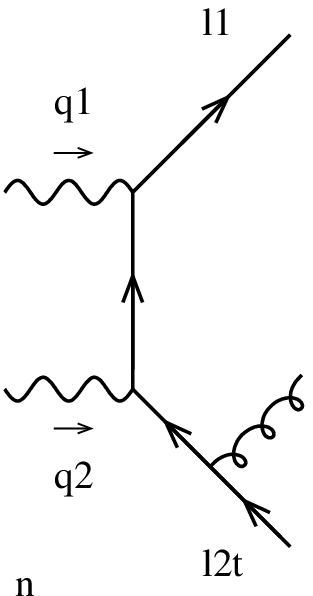,width=\scalings}}&\!+&\!\raisebox{-0.44 \totalheight}{\epsfig{file=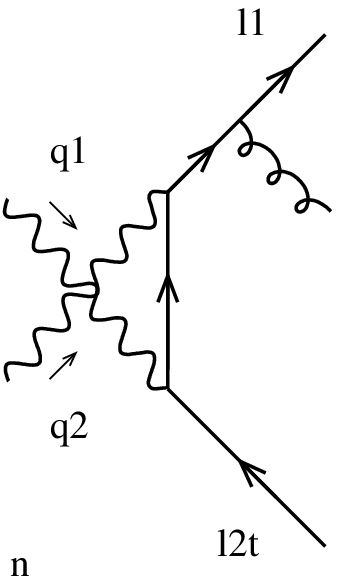,width=\scalingsb}}&\!+&\!\raisebox{-0.44 \totalheight}{\epsfig{file=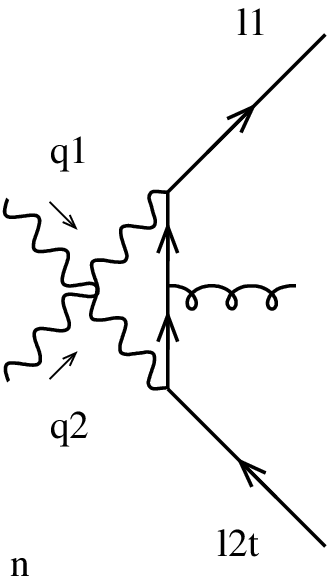,width=\scalingsb}}&\!+&\left.\!\raisebox{-0.44 \totalheight}{\epsfig{file=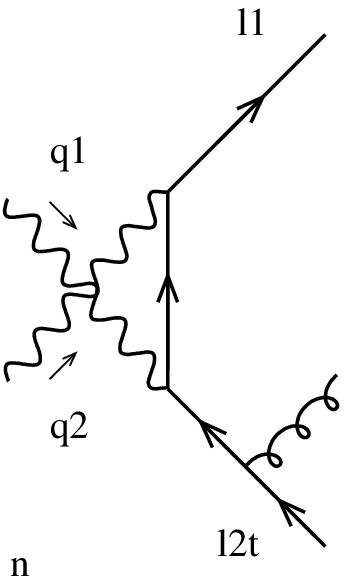,width=\scalingsb}}\right)&\!\!\!\otimes& \!\!\!\raisebox{-0.44 \totalheight}
{\psfrag{l1t}[cc][cc]{\!\!-\lut}
\psfrag{l2}[cc][cc]{\ld}
\psfrag{n}[cc][cc]{}
\epsfig{file=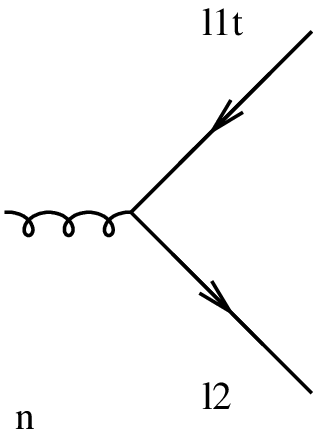,width=\scalings}}\\
& & & & & & & & & & & &\\
(s1) & & (s2) & & (s3) & & (s4) & & (s5) & & (s6) & &\end{array}
&& \\
&& \\
&\hspace{-20cm}+& \\
& &\\
\psfrag{q1}[cc][cc]{$q_1$}
\psfrag{q2}[cc][cc]{$q_2$}
\psfrag{l1}[cc][cc]{\!\lu}
\psfrag{l1t}[cc][cc]{\!-\lut}
\psfrag{l2}[cc][cc]{\ld}
\psfrag{l2t}[cc][cc]{-\ldt}
\psfrag{n}[cc][cc]{}
\begin{array}{ccccccccccccc}
\left(\raisebox{-0.44 \totalheight}{\epsfig{file=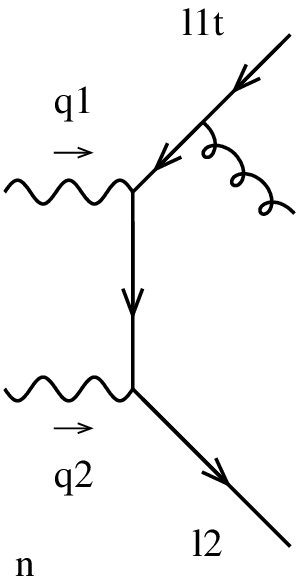,width=\scalings}}\right.&\!+&\!\raisebox{-0.44 \totalheight}{\epsfig{file=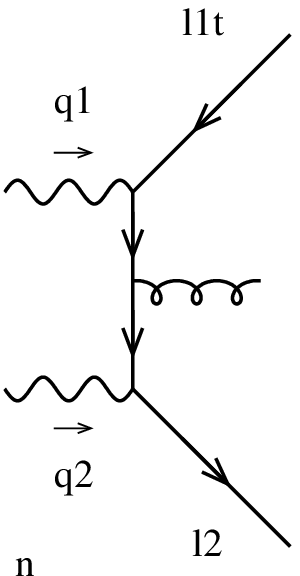,width=\scalings}}&\!+&\!\raisebox{-0.44 \totalheight}{\psfrag{l2t}[cc][cc]{\!\!-\ldt}\epsfig{file=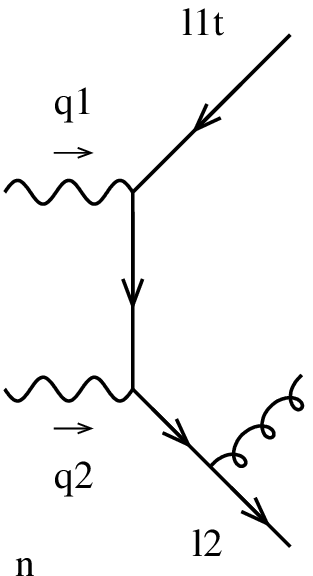,width=\scalings}}&\!+&\!\raisebox{-0.44 \totalheight}{\epsfig{file=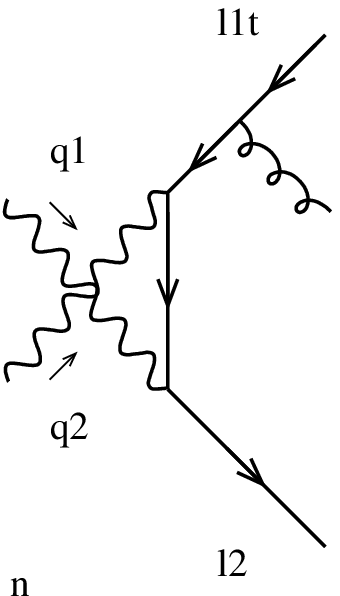,width=\scalingsb}}&\!+&\!\raisebox{-0.44 \totalheight}{\epsfig{file=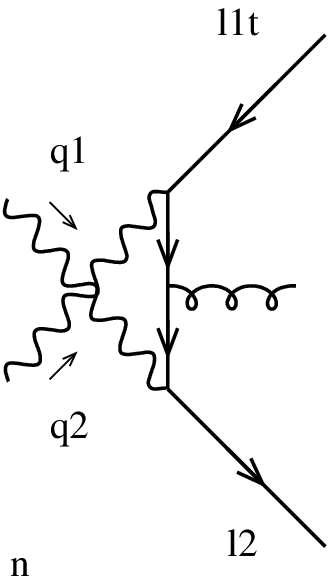,width=\scalingsb}}&\!+&\left.\!\raisebox{-0.44 \totalheight}{\epsfig{file=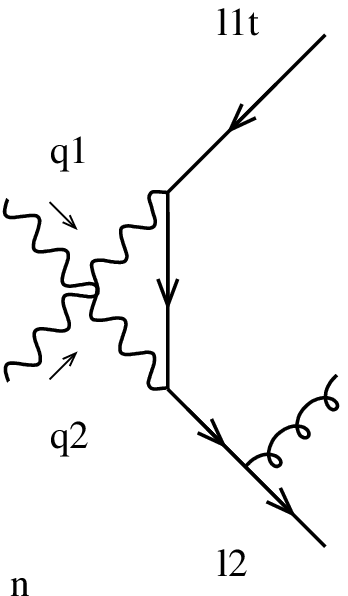,width=\scalingsb}}\right)&\!\!\!\otimes& \!\!\!\raisebox{-0.44 \totalheight}
{\psfrag{l1t}[cc][cc]{\!\!\lu}
\psfrag{l2}[cc][cc]{\!-\ldt}
\psfrag{n}[cc][cc]{}
\epsfig{file=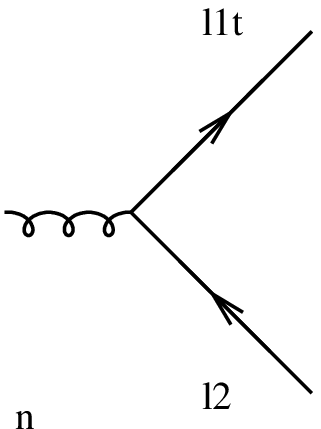,width=\scalings}}\\
& & & & & & & & & & & &\\
(s1') & & (s2') & & (s3') & & (s4') & & (s5') & & (s6') & &\end{array}
\end{eqnarray*}
\caption{Feynman diagrams contributing to $M_H$, in which the virtual photons couple to
a single quark line. \label{sameDiagrams}}
\end{figure}

The Born order contribution to the amplitude (\ref{scatampl}) is calculated
in a similar way as in the classical work of Brodsky-Lepage \cite{BLphysrev24} but in  very different kinematics.
In our case the virtualities of photons supply the hard scale,
and not the transverse momentum transfer. Moreover, we consider the forward kinematics for simplicity.
Relying on the collinear approximation,
the  momenta of the quarks and antiquarks which constitute the rho mesons can be written as
 \bea
\label{defell}
\ell_1 \sim z_1 k_1, &&\quad \ell_2 \sim z_2 k_2 \nonumber \\
\tilde{\ell}_1 \sim \bar{z}_1 k_1, &&\quad \tilde{\ell}_2 \sim \bar{z}_2 k_2 \,.
\eea
The number of possible diagrams is 20. They can be organized into two classes.
The first class corresponds to the diagrams where the two virtual photons couple to two different quark lines.
In this way, one can build 8 different diagrams, as illustrated in Fig.\ref{diffDiagrams}.
The second class of diagrams corresponds to the one where the two virtual photons are coupled to the same quark line, resulting into 12 different contributions, as illustrated in Fig.\ref{sameDiagrams}.
\par
\begin{figure}[htp]
\begin{eqnarray*}
\psfrag{q1}[cc][cc]{$q_1$}
\psfrag{q2}[cc][cc]{$q_2$}
\psfrag{l1}[cc][cc]{\lu}
\psfrag{l1t}[cc][cc]{-\lut}
\psfrag{l2}[cc][cc]{\ld}
\psfrag{l2t}[cc][cc]{\!-\ldt}
\psfrag{n}[cc][cc]{}
\begin{array}{cccccccc}
\raisebox{-0.44 \totalheight}{\epsfig{file=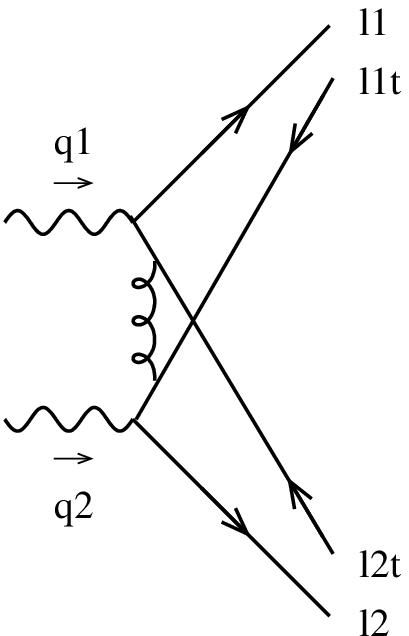,width=\scalingb}}&+&\raisebox{-0.44 \totalheight}{\epsfig{file=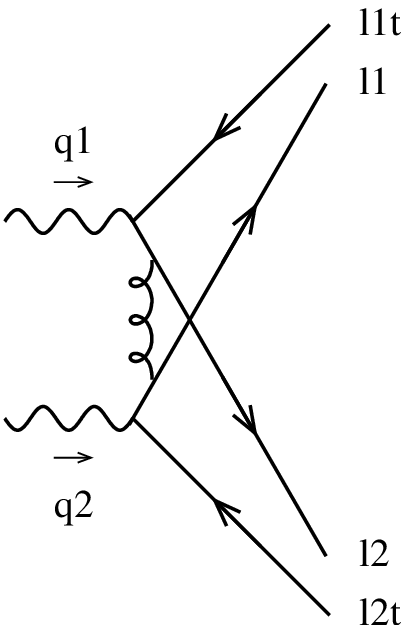,width=\scalingb}} &+&\raisebox{-0.44 \totalheight}{\epsfig{file=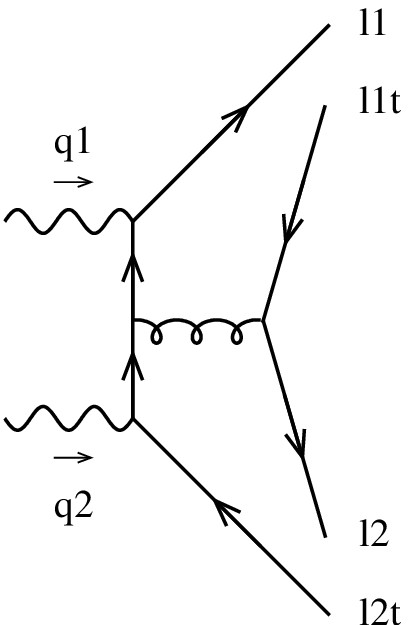,width=\scalingb}}&+&\raisebox{-0.44 \totalheight}{\epsfig{file=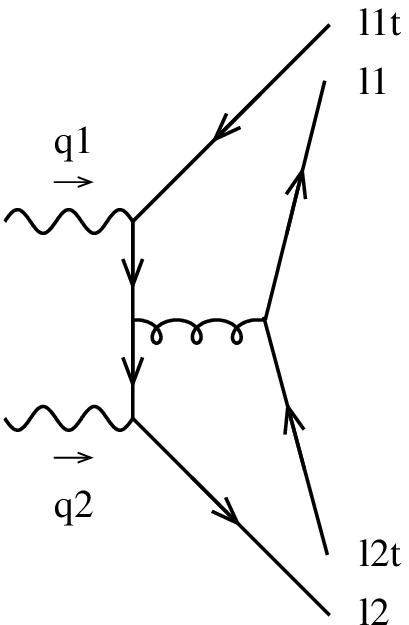,width=\scalingb}}
\end{array}
\end{eqnarray*}
\caption{Feynman diagrams contributing to $M_H$
in the case of
 longitudinally polarized virtual photons. \label{longDiagrams}}
\end{figure}

Let us first focus on the case of longitudinally  polarized photon. Their polarization read
\beq
\epsilon_\parallel(q_1)=\frac{1}{Q_1}q_1 + \frac{2 Q_1}{s}p_2 \quad {\rm and} \quad \epsilon_\parallel(q_2)=\frac{1}{Q_2}q_2 + \frac{2 Q_2}{s}p_1\,.
\ee
Due to the gauge invariance of the total amplitude, when computing
each Feynman diagrams, one can make use of the fact that the longitudinal polarization of photon 1 (resp. 2)
can be effectively considered as proportional to $p_2$ (resp. $p_1$), in accordance with
the structure of the second term of Eq.(\ref{T}).  One then readily sees that in the kinematics we are
investigating, the only diagrams which are non zero in the Feynman gauge
are the diagrams $(1b) \otimes (2a)$, $(3a) \otimes (4b)$ on the one hand, and  $(s2),$ $ (s2')$
on the other hand.
Thus,  only the four diagrams  illustrated in Fig.\ref{longDiagrams} contribute.
\par
In the case of transversally polarized virtual photons, described by the first term in Eq.(\ref{T}), one needs to contract
the two polarization indices through the two dimensional identity tensor $g^T_{\mu\nu}$.
The four non vanishing graphs corresponding to the coupling of the photons to different quark lines  are $(1a) \otimes (2b),$
$(1b) \otimes (2a)$, $(3a) \otimes (4b)$ and $(3b) \otimes (4a).$
When considering the graphs corresponding to the coupling of the photons to the same quark line, the above mentioned contraction kills the graphs $(s2),$ $(s2')$ and $(s5),$ $(s5'),$ that is the graphs where the gluon
is emitted from the quark connecting the two virtual photons. In this second class of diagrams, 8 diagrams thus remain. The whole set of twelve diagrams to be computed is shown in Fig.\ref{transDiagrams}.\renewcommand{\scalingb}{3.2cm}
\renewcommand{\scalingsb}{2.4cm}
\renewcommand{\scalings}{2.1cm}
\begin{figure}[htp]
\psfrag{q1}[cc][cc]{$q_1$}
\psfrag{q2}[cc][cc]{$q_2$}
\psfrag{l1}[cc][cc]{\lu}
\psfrag{l1t}[cc][cc]{-\lut}
\psfrag{l2}[cc][cc]{\ld}
\psfrag{l2t}[cc][cc]{\!-\ldt}
\psfrag{n}[cc][cc]{}
\begin{eqnarray*}
&&\hspace{-.7cm}\begin{array}{cccccccc}
\raisebox{-0.44 \totalheight}{\epsfig{file=al1.eps,width=\scalingb}}&+&\raisebox{-0.44 \totalheight}{\epsfig{file=al2.eps,width=\scalingb}} &+&\raisebox{-0.44 \totalheight}{\epsfig{file=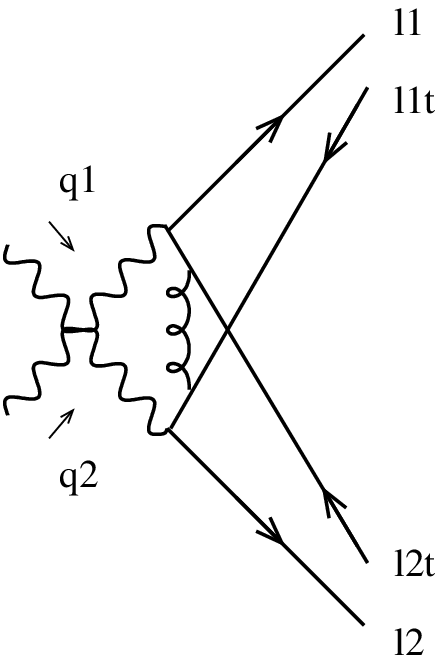,width=\scalingb}}&+&\raisebox{-0.44 \totalheight}{\epsfig{file=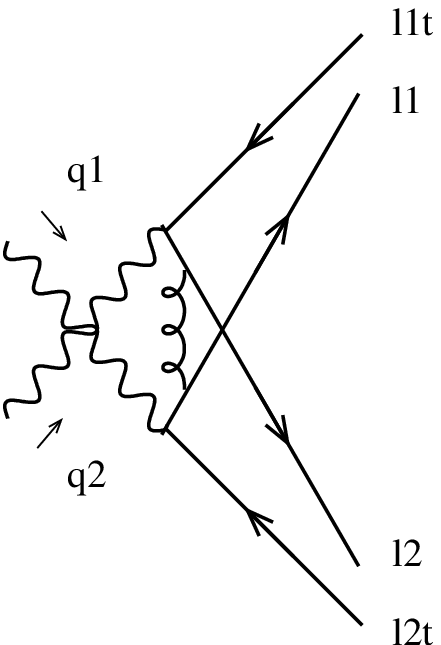,width=\scalingb}}
\end{array}\\
\psfrag{q1}[cc][cc]{$q_1$}
\psfrag{q2}[cc][cc]{$q_2$}
\psfrag{l1}[cc][cc]{\lu}
\psfrag{l1t}[cc][cc]{-\lut}
\psfrag{l2}[cc][cc]{\ld}
\psfrag{l2t}[cc][cc]{\!-\ldt}
\psfrag{n}[cc][cc]{}
&&\\
&&\\
&&\hspace{-.7cm}\begin{array}{cccccccccc}
+&\left(\raisebox{-0.44 \totalheight}{\epsfig{file=d1.eps,width=\scalings}}\right.&\!+&\!\raisebox{-0.44 \totalheight}{\psfrag{l2t}[cc][cc]{\!\!-\ldt}\epsfig{file=d3.eps,width=\scalings}}&\!+&\!\raisebox{-0.44 \totalheight}{\epsfig{file=d4.eps,width=\scalingsb}}&\!+&\left.\!\raisebox{-0.44 \totalheight}{\epsfig{file=d6.eps,width=\scalingsb}}\right)&\!\!\!\otimes& \!\!\!\raisebox{-0.44 \totalheight}
{\psfrag{l1t}[cc][cc]{\!\!-\lut}
\psfrag{l2}[cc][cc]{\ld}
\psfrag{n}[cc][cc]{}
\epsfig{file=splitting1.eps,width=\scalings}}\\
&  & & & & & & & &\\
&(s1)  & & (s3) & & (s4) & & (s6) & &\end{array}\\
&& \\
&& \\
& &\\
\psfrag{q1}[cc][cc]{$q_1$}
\psfrag{q2}[cc][cc]{$q_2$}
\psfrag{l1}[cc][cc]{\!\lu}
\psfrag{l1t}[cc][cc]{\!-\lut}
\psfrag{l2}[cc][cc]{\ld}
\psfrag{l2t}[cc][cc]{-\ldt}
\psfrag{n}[cc][cc]{}
&&\hspace{-.7cm}\begin{array}{cccccccccc}
+&\left(\raisebox{-0.44 \totalheight}{\epsfig{file=d7.eps,width=\scalings}}\right.&\!+&\!\raisebox{-0.44 \totalheight}{\psfrag{l2t}[cc][cc]{\!\!-\ldt}\epsfig{file=d9.eps,width=\scalings}}&\!+&\!\raisebox{-0.44 \totalheight}{\epsfig{file=d10.eps,width=\scalingsb}}&\!+&\left.\!\raisebox{-0.44 \totalheight}{\epsfig{file=d12.eps,width=\scalingsb}}\right)&\!\!\!\otimes& \!\!\!\raisebox{-0.44 \totalheight}
{\psfrag{l1t}[cc][cc]{\!\!\lu}
\psfrag{l2}[cc][cc]{\!-\ldt}
\psfrag{n}[cc][cc]{}
\epsfig{file=splitting2.eps,width=\scalings}}\\
&& & & & & & & & \\
&(s1') & & (s3') & & (s4') & & (s6') & &\end{array}
\end{eqnarray*}
\caption{Feynman diagrams contributing to $M_H$
in the case of
 transversally polarized virtual photons. \label{transDiagrams}}
\end{figure}
\par
It should be noted that in the peculiar kinematics which is chosen here,  the only diagrams which contribute both to the longitudinal and to the transverse virtual photon cases are $(1b) \otimes (2a)$, $(3a) \otimes (4b),$ that is the two first diagrams of Fig.\ref{longDiagrams} and Fig.\ref{transDiagrams}.

The scalar components of the scattering amplitude (\ref{T}) read :
\begin{eqnarray}
\label{Ttr}
T^{\alpha\, \beta}g_{T\,\alpha \,\beta}&=& -\frac{e^2(Q_u^2 +Q_d^2)\,g^2\,C_F\,f_\rho^2}{4\,N_c\,s}
\int\limits_0^1\,dz_1\,dz_2\,\phi(z_1)\,\phi(z_2)
 \\
&&\hspace{-2cm}\times \left\{2\left(1-\frac{Q^2_2}{s}\right)\left(1-\frac{Q^2_1}{s}\right)\left[
\frac{1}{(z_2+\bar z_2\frac{Q_1^2}{s})^2(z_1+\bar z_1\frac{Q_2^2}{s} )^2} +
\frac{1}{(\bar z_2+ z_2\frac{Q_1^2}{s})^2(\bar z_1+ z_1\frac{Q_2^2}{s} )^2}
 \right] \; + \right.
\nonumber \\
 && \hspace{-3cm}\left.
\left(\frac{1}{\bar z_2\,z_1}- \frac{1}{\bar z_1\,z_2}  \right)
\left[ \frac{1}{1-\frac{Q^2_2}{s}}\left( \frac{1}{\bar z_2+ z_2\frac{Q_1^2}{s}}
                                      -\frac{1}{ z_2+ \bar z_2\frac{Q_1^2}{s}}   \right)
 - \frac{1}{1-\frac{Q^2_1}{s}}\left(\frac{1}{\bar z_1+ z_1\frac{Q_2^2}{s}}                                    - \frac{1}{ z_1+ \bar z_1\frac{Q_2^2}{s}}   \right) \right]
\right\} \nonumber \\
\label{Tlong}
T^{\alpha\, \beta}p_{2\,\alpha} \,p_{1\,\beta}&=& -\frac{s^2 f_\rho^2 C_F e^2 g^2(Q_u^2 +Q_d^2)}{8N_{c}Q_1^2 Q_2^2}
\int\limits_0^1\,dz_1\,dz_2\,\phi(z_1)\,\phi(z_2)
 \\
&&\times \left\{\frac{(1-\frac{Q^2_1}{s})(1-\frac{Q^2_2}{s})}{(z_1+\bar z_1 \frac{Q^2_2}{s})(z_2+\bar z_2\frac{Q^2_1}{s})}
+ \frac{(1-\frac{Q^2_1}{s})(1-\frac{Q^2_2}{s})}{(\bar z_1+ z_1 \frac{Q^2_2}{s})(\bar z_2+ z_2\frac{Q^2_1}{s})}
+ \frac{1}{z_2 \bar z_1} + \frac{1}{z_1 \bar z_2} \right\} \nonumber\;,
\end{eqnarray}

where $Q_u=2/3$ ($Q_d=-1/3$) denote the charge of the quark $u$ ($d$),
$C_F=(N_c^2-1)/(2 N_c)$ and $N_c=3.$

The above results are obtained for  arbitrary values of the photon virtualities $Q_i$.
A closer look into formulas (\ref{scatampl}-\ref{Tlong}) leads to the conclusion that all integrals over
quarks momentum fractions  $z_i$ are convergent due to the non-zero values of  $Q_i$.  Further inspection of the amplitudes
 (\ref{scatampl}-\ref{Tlong}) reveals that the transverse photon part has a behaviour like $1/W^2$
while the longitudinal photon part behaves like $1/(Q_1 Q_2).$
This shows that their relative magnitude is different in each
   region (1) and (2) discussed in Sec. 1.

\section{$\gamma_T^* \gamma_T^* \to \rho^0_L \rho^0_L$ in the generalized Bjorken limit}
\label{secgda}

\begin{figure}[htb]
\psfrag{r1}[cc][cc]{$\quad\rho(k_1)$}
\psfrag{r2}[cc][cc]{$\quad\rho(k_2)$}
\psfrag{p1}[cc][cc]{$\slashchar{p}_1$}
\psfrag{p2}[cc][cc]{$\slashchar{p}_2$}
\psfrag{p}[cc][cc]{$\qquad\slashchar{P}\qquad \slashchar{n}$}
\psfrag{n}[cc][cc]{}
\psfrag{q1}[cc][cc]{$q_1$}
\psfrag{q2}[cc][cc]{$q_2$}
\psfrag{GDA}[cc][cc]{$GDA_H$}
\psfrag{Da}[cc][cc]{DA}
\psfrag{HDA}[cc][cc]{$M_H$}
\psfrag{M}[cc][cc]{$M$}
\psfrag{Th}[cc][cc]{$T_H$}
\centerline{\scalebox{1}
{$\begin{array}{cccccc}
\!\!\raisebox{-0.44\totalheight}{\epsfig{file=HDA.eps,width=\scalingD}}& \begin{array}{c}
\raisebox{0.51 \totalheight}
{\psfrag{r}[cc][cc]{$\quad\rho(k_1)$}
\psfrag{pf}[cc][cc]{$\slashchar{p}_2$}
\epsfig{file=DA.eps,width=\scalingA}}\\
\raisebox{0. \totalheight}
{\psfrag{r}[cc][cc]{$\quad\rho(k_2)$}
\psfrag{pf}[cc][cc]{$\slashchar{p}_1$}
\epsfig{file=DA.eps,width=\scalingA}}
\end{array}&=& \!
\raisebox{-0.44 \totalheight}
{\psfrag{r}[cc][cc]{$\quad\rho(k_1)$}
\psfrag{pf}[cc][cc]{$\slashchar{p}_2$}
\epsfig{file=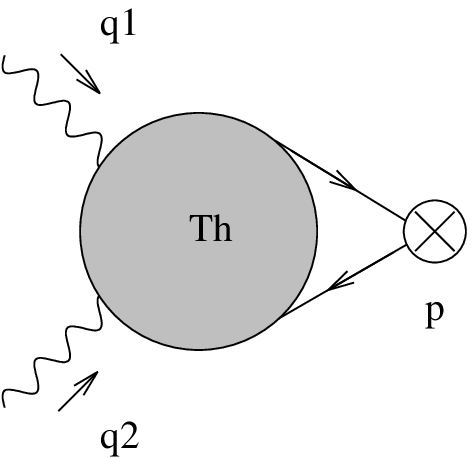,width=\scalingB}}&
\raisebox{-0.43\totalheight}{\epsfig{file=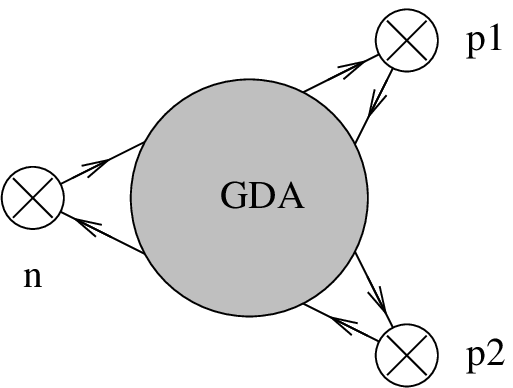,width=\scalingC}}& \begin{array}{c}
\raisebox{0.51 \totalheight}
{\psfrag{r}[cc][cc]{$\quad\rho(k_1)$}
\psfrag{pf}[cc][cc]{$\slashchar{p}_2$}
\epsfig{file=DA.eps,width=\scalingA}}\\
\raisebox{-0.0 \totalheight}
{\psfrag{r}[cc][cc]{$\quad\rho(k_2)$}
\psfrag{pf}[cc][cc]{$\slashchar{p}_1$}
\epsfig{file=DA.eps,width=\scalingA}}
\end{array}
\end{array}
$}}
\caption{Factorisation of the  amplitude in terms of a GDA. \label{FactGDA}}
\end{figure}

Let us first focus on the region where the scattering energy $W$ is
small in comparison with the highest photon virtuality $Q_1$ \be
\label{scaling} \frac{W^2}{Q_1^2}=
\frac{s}{Q_1^2}\left(1-\frac{Q_1^2}{s}
\right)\left(1-\frac{Q_2^2}{s}  \right) \approx 1-\frac{Q_1^2}{s}\ll
1\;, \ee which leads to the kinematical conditions very close to the
ones considered in Ref.\cite{DM,GDA} for the  description of
$\gamma^* \gamma \to \pi \pi$ near threshold. In Ref.\cite{DM, GDA}
it was shown that with initial transversally polarized photons, the
scattering amplitude factorizes at leading twist as the convolution
of a perturbatively calculable coefficient function  and a
generalized distribution amplitude (GDA). We will recover a similar
type of factorization with a GDA of the expression (\ref{Ttr}) also
in the case of our process (\ref{proces}), as illustrated in
Fig.\ref{FactGDA},
provided $Q_1$ and $Q_2$ are not parametrically close, i.e.
 \beq
 \label{XY} 1-\frac{Q_1^2}{s} \ll 1-\frac{Q_2^2}{s}\,.
  \ee

Indeed, in the scaling limit (\ref{scaling}) with (\ref{XY}), the
contribution of the four cats-ears diagrams (i.e. the first line of
Fig.\ref{transDiagrams}), which corresponds to the second line of
(\ref{Ttr}), is subdominant and the dominant contribution is given
by the last term in (\ref{Ttr}), i.e.
\begin{eqnarray}
\label{Ttrscaling}
&&T^{\alpha\, \beta}g_{T\,\alpha \,\beta} \approx \frac{e^2(Q_u^2 +Q_d^2)\,g^2\,C_F\,f_\rho^2}{4\,N_c\,W^2}
\int\limits_0^1\,dz_1\,dz_2\,\phi(z_1)\,\phi(z_2)
 \\
 &&
\left(\frac{1}{\bar z_2\,z_1}- \frac{1}{\bar z_1\,z_2}  \right)
\left(\frac{1}{\bar z_1+ z_1\frac{Q_2^2}{s}}
 - \frac{1}{ z_1+ \bar z_1\frac{Q_2^2}{s}}   \right)\;.
 \nonumber
\end{eqnarray}
The virtuality $Q_2$ plays in the expression (\ref{Ttrscaling}) the role of a regulator
of the end-point singularities (compare with \cite{GDA}).
Before recovering each factor of the factorized  equation (\ref{Ttrscaling}) by a direct calculation, let us discuss the physics which is behind.
In Fig.\ref{transDiagrams}, the diagrams which contribute in the scaling region (\ref{scaling}) with (\ref{XY})
 are $(s3),$  $(s3')$ and  $(s6),$  $(s6').$
The sum of these four diagrams factorizes into a hard part convoluted with a soft (but still perturbative) part. Indeed, the typical virtuality
of the hard quark connecting the two virtual photons is of the order of
$Q_1^2$ for graphs  $(s3),$  $(s3')$ and $s$ for graphs  $(s6),$  $(s6'),$
while the virtuality of the quark connecting the emitted gluon
to the virtual photon is of the order of $W^2,$ which is negligible
in the scaling region with respect to $Q_1^2$ and $s.$

Let us now recall the definition of the leading twist GDA for $\rho^0_L$ pair. We introduce the
vector $P=k_1+k_2 \approx p_1$, whereas the field coordinates  are the ray-vectors along
the light-cone direction $n^\mu=p_2^\mu/(p_1.p_2)$.
 In our kinematics the usual variable $\zeta = (k_1n)/(Pn)$ characterizing the GDA equals $\zeta \approx 1$.
Thus we define the GDA of the $\rho^0_L$ pair $\Phi_q(z,\zeta,W^2)$  by
the formula
\begin{eqnarray}
\label{GDA}
&&\langle \rho^0_L(k_1)\,\rho^0_L(k_2)| \bar q(-\alpha \;n/2)\,\slashchar{n}\,
\left[\int\limits_{-\frac{\alpha}{2}}^{\frac{\alpha}{2}}\,dy \, n_\nu\, A^\nu (y) \right]
q(\alpha \;n/2)|0\rangle
\nonumber \\
&&=
 \int\limits_0^1\,dz\,e^{-i(2z-1)\alpha (nP)/2}\Phi^{\rho_L\rho_L}(z,\zeta,W^2)\;,\;\;\;\; q=u,d.
\end{eqnarray}

\begin{figure}[htb]
\psfrag{r1}[cc][cc]{$\quad\rho(k_1)$}
\psfrag{r2}[cc][cc]{$\quad\rho(k_2)$}
\psfrag{p1}[cc][cc]{$\slashchar{p}_1$}
\psfrag{p2}[cc][cc]{$\slashchar{p}_1$}
\psfrag{q1}[cc][cc]{$q_1$}
\psfrag{q2}[cc][cc]{$q_2$}
\psfrag{Da}[cc][cc]{DA}
\psfrag{HDA}[cc][cc]{$M_H$}
\psfrag{M}[cc][cc]{$M$}
\centerline{\scalebox{1}
{
$\begin{array}{ccccc}
\raisebox{-0.44 \totalheight}{\epsfig{file=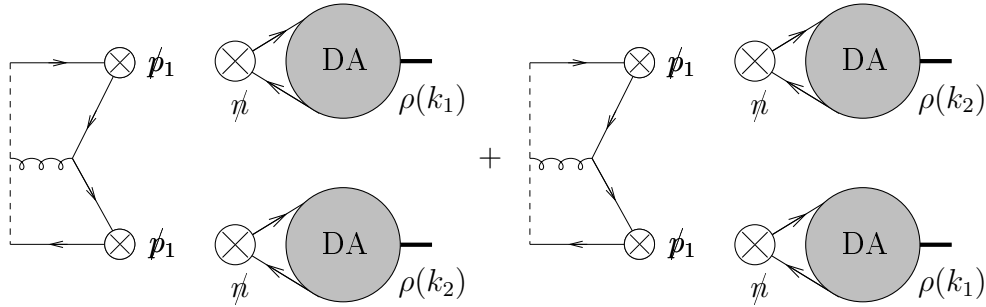,width=\scalingm}}
& \begin{array}{c}
\raisebox{0.45 \totalheight}
{\psfrag{r}[cc][cc]{$\quad\rho(k_1)$}
\psfrag{pf}[cc][cc]{$\slashchar{n}$}
\epsfig{file=DA.eps,width=\scaling}}\\
\raisebox{-0. \totalheight}
{\psfrag{r}[cc][cc]{$\quad\rho(k_2)$}
\psfrag{pf}[cc][cc]{$\slashchar{n}$}
\epsfig{file=DA.eps,width=\scaling}}
\end{array}
&+&
\raisebox{-0.44\totalheight}{\epsfig{file=GDAW1.eps,width=\scalingm}}
& \begin{array}{c}
\raisebox{0.45 \totalheight}
{\psfrag{r}[cc][cc]{$\quad\rho(k_2)$}
\psfrag{pf}[cc][cc]{$\slashchar{n}$}
\epsfig{file=DA.eps,width=\scaling}}\\
\raisebox{-0. \totalheight}
{\psfrag{r}[cc][cc]{$\quad\rho(k_1)$}
\psfrag{pf}[cc][cc]{$\slashchar{n}$}
\epsfig{file=DA.eps,width=\scaling}}
\end{array}
\end{array}
$}}
\caption{Wilson line contribution to the GDA, expressed as a convolution of a hard part with the DAs. \label{WilsonQCD}}
\end{figure}

Now we calculate the GDA  $\Phi_q(z,\zeta,W^2)$ in the Born order of the perturbation theory.
First we show that the gluonic Wilson line does not give a contribution in our kinematics. For that
 expand the Wilson line and the S-matrix operator with quark-gluon interaction in (\ref{GDA})
at the first order in $g$. We obtain (up to an irrelevant multiplicative factor)
\be
n_\mu g^2\int d^4v \, \langle \rho^0_L(k_1)\,\rho^0_L(k_2)|
T[\bar q(-\alpha \;n/2)\,\gamma^\mu
\left[\int\limits_{-\frac{\alpha}{2}}^{\frac{\alpha}{2}}\,dy\, n_\nu\, A^\nu (y) \right] q(\alpha \;n/2) \,
\bar q(v) \slashchar{A}(v) q(v) ]|0\rangle
\ee
After applying the Fierz identity and ordering quark operators into two non-local correlators defining DA of $\rho-$mesons
we obtain
\be
n_\mu \langle \rho^0_L(k_1)| \bar q(-\alpha n/2)\,\gamma^\delta \,q(v)|0\rangle
\langle \rho^0_L(k_2)| \bar q(v)\,\gamma^\sigma \,q(\alpha n/2)|0\rangle Tr[\gamma^\delta \gamma^\mu \gamma^\sigma
\gamma^\beta] n_{\beta} + (k_1 \leftrightarrow k_2)\;,
\ee
where we omitted the gluon propagator coming from the contraction of the two gauge fields.
The original Wilson line results at this order
in the presence of the  vector $n_{\beta} $ in the above expression.
The only nonvanishing contribution at leading twist  takes the form
\be
\langle \rho^0_L(k_1)| \bar q(-\alpha n/2)\,\slashchar{ n} \,q(v)|0\rangle
\langle \rho^0_L(k_2)| \bar q(v)\,\slashchar{ n} \,q(\alpha n/2)|0\rangle Tr[\slashchar{p}_1 \slashchar{n} \slashchar{p}_1 \slashchar{n}]
+ (k_1 \leftrightarrow k_2)\;.
\ee
This is illustrated in Fig.\ref{WilsonQCD}.
\par
It equals zero since in our kinematics one of the matrix elements defining DA of $\rho-$meson vanishes.

\begin{figure}[htb]
\psfrag{p1}[cc][cc]{$\slashchar{p}_1$}
\psfrag{p2}[cc][cc]{$\slashchar{p}_2$}
\psfrag{n}[cc][cc]{$\slashchar{n}$}
\centerline{\scalebox{1.2}
{
$\begin{array}{ccc}
\raisebox{-0.44 \totalheight}{\epsfig{file=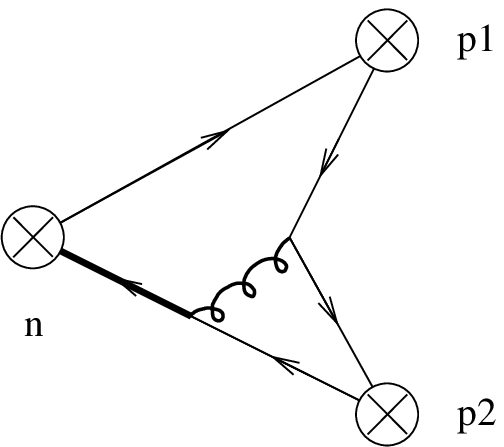,width=\scaling}}
&+&
\raisebox{-0.44\totalheight}{\epsfig{file=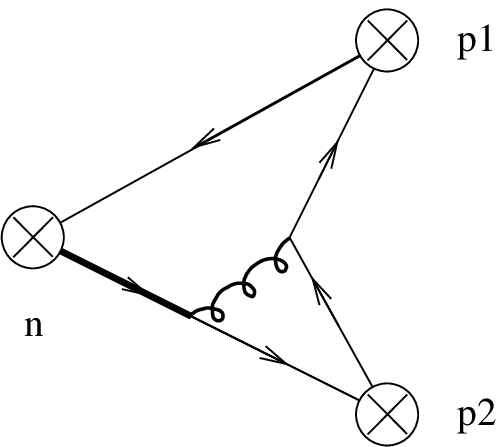,width=\scaling}}
\end{array}
$}}
\caption{Non vanishing contributions to the hard part of the GDA at Born order. The quark and gluon bold lines
correspond to propagators
.  \label{devGDA}}
\end{figure}

The  remaining contributions to
the correlator in (\ref{GDA}) at order $g^2$  are illustrated in Fig.\ref{devGDA}.

\begin{figure}[htb]
\psfrag{p}[cc][cc]{$\slashchar{P}$}
\psfrag{q1}[cc][cc]{$q_1$}
\psfrag{q2}[cc][cc]{$q_2$}
\psfrag{Th}[cc][cc]{$T_H$}
\psfrag{n}[cc][cc]{$\slashchar{n}$}
\centerline{\scalebox{1.2}
{
$\begin{array}{ccccc}
\raisebox{-0.46 \totalheight}{\epsfig{file=THT.eps,width=\scaling}}
&=&
\raisebox{-0.46 \totalheight}{\epsfig{file=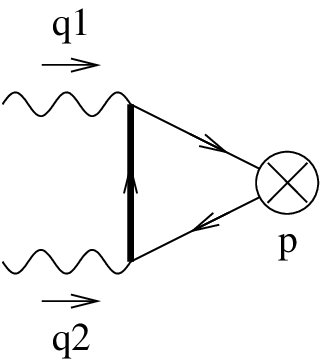,width=\scaling}}
&+&
\raisebox{-0.46\totalheight}{\epsfig{file=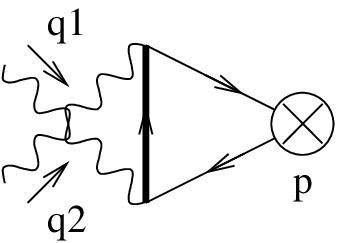,width=\scaling}}
\end{array}
$}}
\caption{Expansion of the hard part $T_H$ at $g^2$ order.  The bold lines correspond to quark propagators. \label{devTHT}}
\end{figure}

It leads to the result
\be
\label{Phi}
\Phi^{\rho_L\rho_L}(z,\zeta\approx 1,W^2)
= -\frac{f_\rho^2\,g^2\,C_F}{2\,N_c\,W^2}\int\limits_0^1\,dz_2\,\phi(z)\,\phi(z_2)
\left[\frac{1}{z\bar z_2} -\frac{1}{\bar z z_2} \right]\;.
\ee
The hard part $T_H$ of the amplitude corresponds to the diagrams shown
in Fig.\ref{devTHT}.
In the case of a quark of a given flavour it equals
\begin{equation}
\label{hardGDA}
T_H(z) = -4\,e^2\,N_c\,Q_q^2\,\left( \frac{1}{\bar z +z\,\frac{Q_2^2}{s}} -
   \frac{1}{z + \bar z\,\frac{Q_2^2}{s}} \right)
\end{equation}
The Eqs.~(\ref{Phi}, \ref{hardGDA}) taken together with the flavour
structure of $\rho^0$ permit to write (\ref{Ttrscaling}) in the form
\be
\label{GDAfinal}
T^{\alpha\, \beta}g_{T\,\alpha \,\beta}=
\frac{ e^2}{2}\left(Q_u^2 +Q_d^2  \right)\int\limits_0^1 \,dz\,\left(\frac{1}{\bar z +z \frac{Q_2^2}{s}}
- \frac{1}{ z +\bar z \frac{Q_2^2}{s}}  \right)\Phi^{\rho_L\rho_L}(z,\zeta\approx 1,W^2)\;,
\ee
which shows the factorization of $T^{\alpha\, \beta}g_{T\,\alpha \,\beta}$
 into the hard part and the GDA.
The Eq.~(\ref{GDAfinal})
 is  the limiting case for $\zeta \to 1$
of the original equation derived by D.~M{\"u}ller et al \cite{DM}.

\section{$\gamma_L^* \gamma_L^* \to \rho^0_L \rho^0_L$
in the generalized Bjorken limit}
\label{sectda}
To analyse the case of longitudinally polarized photon scattering,
let us now turn to another interesting limiting case where
 \beq
\label{region} Q_{1}^2 ~~ >>~~ Q_{2}^2 \;.
\ee
In this
regime, it  has been advocated \cite{TDA} that the amplitude with
initial longitudinally polarized  photons, should factorize as the
convolution of a perturbatively calculable coefficient function  and
a $\gamma \to \rho$ transition distribution amplitude (TDA) defined
from the non-local quark correlator
  \beq
\int \frac{dz^-}{2\pi} e^{-ixP^+z^-}< \rho(p_{2}) | \bar q(-z^-/2) \gamma^+ q(z^-/2)) | \;\gamma(q_{2})>\;,
\ee
 which shares many properties (including the QCD evolution equations) with the generalized parton
 distributions \cite{DM,GPD} succesfully introduced to describe deeply virtual Compton scattering.

The factorization properties of the scattering amplitude in this
domain are conventionally described by using the now standard
notations of GPDs. For that we rewrite the momenta of the particles
involved in the process as
\begin{eqnarray}
\label{GPDkin}
&&q_1=\frac{1}{1+\xi}\,n_1 -2\xi\,n_2\;,\;\;\;\;\;k_1=\frac{1-\frac{Q_2^2}{s}}{1+\xi}\,n_1
\nonumber \\
&& q_2=-\frac{Q_2^2}{(1+\xi)s}\,n_1 + (1+\xi)\,n_2\;,\;\;\;\;k_2=(1-\xi)\,n_2\;,
\end{eqnarray}
where $\xi$ is the skewedness parameter which equals $\xi= Q_1^2/(2s - Q_1^2)$ (see also the formula (\ref{sW}) which relates $s$
with the total scattering energy $W$) and the new $n_i$ Sudakov light-cone
vectors are related to the $p_i$'s as
\begin{equation}
\label{np} p_1=\frac{1}{1+\xi}\,n_1\;,\;\;\;\;\;p_2=(1+\xi)\,n_2\;,
\end{equation}
with
 $p_1 \cdot p_2=n_1 \cdot n_2=s/2$. We also introduce the
average ``target'' momentum $P$ and the momentum transfer $\Delta$
\begin{equation}
\label{PDelta}
P=\frac{1}{2}(q_2 +k_2)\;, \;\;\;\;\;\;\Delta=k_2-q_2\;.
\end{equation}
We still restrict our study to the strictly forward case with $t=t_{min} =-2\xi Q_2^2/(1+\xi)$.

In the region defined by Eq.~(\ref{region})
we can put $Q_2=0$ inside $\{...\}$
in the expression in Eq.~(\ref{Tlong}), which results, in this approximation,  in
the formula:
\begin{eqnarray}
\label{Tlongsimpl}
&&\int\limits_0^1\,dz_1\,dz_2\,\phi(z_1)\,\phi(z_2)\left\{....\right\}
 \\
&&=
\int\limits_0^1\,dz_1\,dz_2\,\phi(z_1)\,\phi(z_2)
\left\{
 \frac{(1-\xi)}{\bar z_1[\bar z_2(1+\xi) +2\xi\, z_2]}
+\frac{(1-\xi)}{z_1[z_2(1+\xi) +2\xi\,\bar z_2]} +\frac{1}{z_2\,\bar z_1} + \frac{1}{z_1\,\bar z_2} \right\}\;.
\nonumber
\end{eqnarray}
The terms in the $\{\cdots \}$ are ordered in accordance with
diagrams shown in Fig.\ref{longDiagrams}.

\begin{figure}[htb]
\psfrag{q1}[cc][cc]{$q_1$}
\psfrag{q2}[cc][cc]{$q_2$}
\psfrag{p1}[cc][cc]{$\slashchar{p}_1$}
\psfrag{p2}[cc][cc]{$\slashchar{p}_2$}
\psfrag{n}[cc][cc]{$\slashchar{p}_1$}
\psfrag{p}[cc][cc]{$\slashchar{p}_2$}
\psfrag{Tda}[cc][cc]{$TDA_H$}
\psfrag{Th}[cc][cc]{$T_H$}
\psfrag{Da}[cc][cc]{DA}
\centerline{\scalebox{1.2}
{
$\begin{array}{c}
\begin{array}{cc}
\raisebox{-0.44 \totalheight}{\epsfig{file=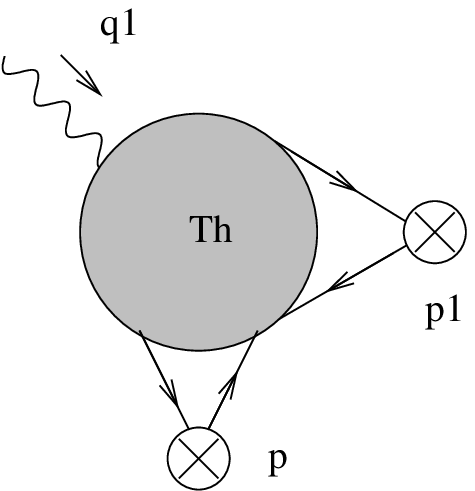,width=\scaling}}&
\raisebox{-0.25\totalheight}
{\psfrag{r}[cc][cc]{$\qquad \rho(k_1)$}
\psfrag{pf}[cc][cc]{\raisebox{-1.05 \totalheight}{$\slashchar{p}_2$}}\epsfig{file=DA.eps,width=\scalingm}}
\end{array}\\
\\
\begin{array}{cc}
\raisebox{-0.44 \totalheight}{\epsfig{file=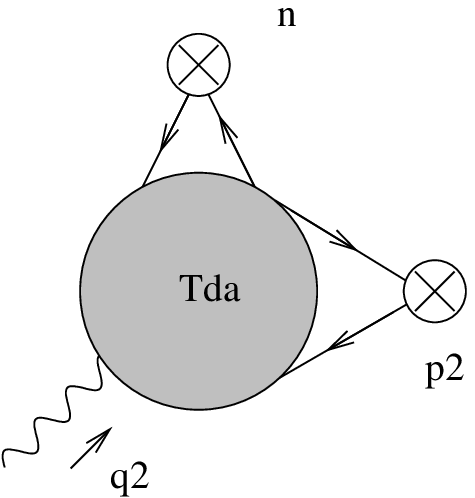,width=\scaling}}&
\raisebox{-0.54\totalheight}
{\psfrag{r}[cc][cc]{$\qquad \rho(k_2)$}
\psfrag{pf}[cc][cc]{\raisebox{-1.05 \totalheight}{$\slashchar{p}_1$}}\epsfig{file=DA.eps,width=\scalingm}}
\end{array}
\end{array}
$}}
\caption{Factorization of the amplitude in terms of a TDA.  \label{FactTDA}}
\end{figure}

Now our aim is to rewrite Eq.~(\ref{Tlongsimpl}) in a form
corresponding to the QCD factorization with a TDA, as illustrated in
Fig.\ref{FactTDA}.

For that we look more closely into diagrams contributing to each of
the four terms in (\ref{Tlongsimpl}) from the point of view of such
a factorization. For example, the second term in (\ref{Tlongsimpl}),
corresponding to the second diagram shown in Fig.\ref{longDiagrams},
suggests the introduction of the new variable $x$, defined through
$z_2=(x-\xi)/(1-\xi)$, with $x \in [\xi,1]$, which results in the
equality
\begin{equation}
\label{1term}
\int\limits_0^1 \,dz_2\,\frac{\phi(z_2)(1-\xi)}{z_1[z_2(1+\xi)+2\xi\,\bar z_2]}=
\int\limits_\xi^1\, dx\,\frac{\phi\left(\frac{x-\xi}{1-\xi}    \right)}{z_1(x+\xi)}\;,
\end{equation}
i.e. which corresponds to a part of the contribution from the DGLAP integration region of
the amplitude with a factorized
TDA. A similar analysis of all
remaining terms permits to represent Eq.~(\ref{Tlongsimpl}) in the form
\begin{eqnarray}
\label{4terms}
&&\int\limits_0^1\,dz_1\,dz_2\,\phi(z_1)\,\phi(z_2)\left\{....\right\}
 \\
&&=
\int\limits_{-1}^1\,dx\,
\int\limits_0^1\,dz_1\,\phi(z_1)\,\left( \frac{1}{\bar z_1(x-\xi)}+ \frac{1}{z_1(x+\xi)}  \right)
\nonumber \\
&&\times \, \left[\Theta(1\ge x \ge \xi) \phi\left(\frac{x-\xi}{1-\xi} \right) -
\Theta(-\xi \ge x \ge -1) \phi\left( \frac{1+x}{1-\xi} \right)
 \right]\;, \nonumber
\end{eqnarray}
with the step fuction $\Theta(a\ge x \ge b ) = \Theta(a-x)\Theta(x-b)$. This factorized expression suggests
the identification of
\begin{equation}
\label{cfunction}
\int\limits_0^1\,dz_1\,\phi(z_1)\,\left( \frac{1}{\bar z_1(x-\xi)}+ \frac{1}{z_1(x+\xi)}  \right)
\end{equation}
as  the coefficient function $T_H$ (up to a multiplicative factor), and of
\begin{equation}
\label{tda}
T(x,\xi,t_{min})\equiv N_c \left[\Theta(1\ge x \ge \xi)\, \phi\left(\frac{x-\xi}{1-\xi} \right) -
\Theta(-\xi \ge x \ge -1) \, \phi\left( \frac{1+x}{1-\xi} \right) \right]
\end{equation}
as the $\gamma^*_L \to \rho_L$ TDA.

\begin{figure}[htb]
\psfrag{p}[cc][cc]{$\,\, \slashchar{p}_2$}
\psfrag{p1}[cc][cc]{$\,\, \slashchar{p}_1$}
\psfrag{p2}[cc][cc]{$\,\, \slashchar{p}_2$}
\psfrag{q1}[cc][cc]{$q_1$}
\psfrag{Th}[cc][cc]{$T_H$}
\psfrag{n}[cc][cc]{$\slashchar{n}$}
\centerline{\scalebox{1.}
{
$
\begin{array}{ccccc}
\raisebox{-0.46 \totalheight}{\epsfig{file=THL.eps,width=\scalingl}}
&\, =&
\raisebox{-0.46 \totalheight}{\epsfig{file=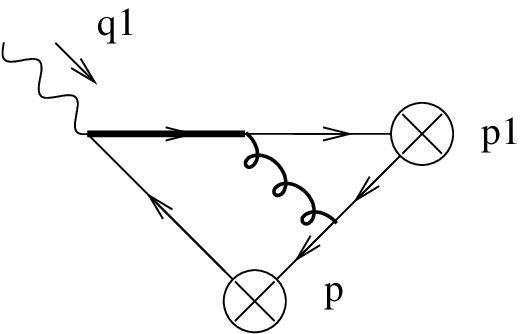,width=\scalingl}}
&\,+&
\raisebox{-0.46 \totalheight}{\epsfig{file=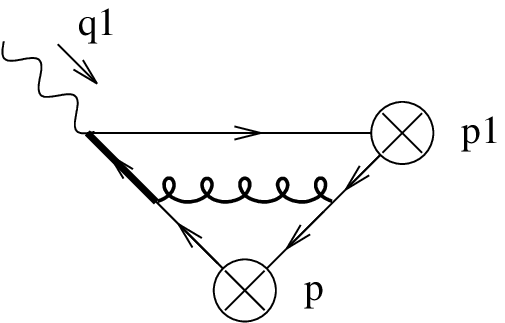,width=\scalingl}}
\\
&\,+&
\raisebox{-0.46\totalheight}{\epsfig{file=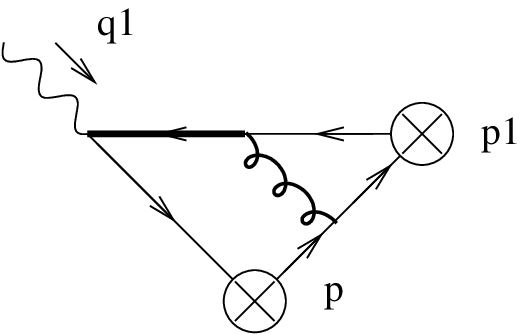,width=\scalingl}}
&\,+&
\raisebox{-0.46\totalheight}{\epsfig{file=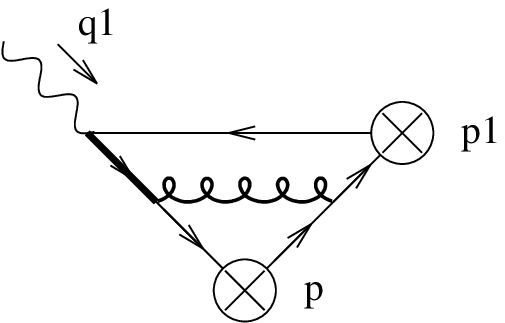,width=\scalingl}}
\end{array}
$}}
\caption{The hard part $T_H$ at  order $g^2$. \label{devTHL}}
\end{figure}

To justify this interpretation we start from
the hard part $T_H(z_1,x)$ of the scattering amplitude, which appears in Fig.\ref{FactTDA},
as illustrated at order $g^2$ in Fig.\ref{devTHL}.
It equals, for a meson built from a quark with a single flavour,
\begin{eqnarray}
\label{hardpart}
T_H(z_1,x)&=&-i\,f_\rho\,g^2\,e\,Q_q\,\frac{C_F\,\phi(z_1)}{2\,N_c\,Q_1^2}
\epsilon^\mu(q_1)\left(2\xi\,n_{2\,\mu}  +\frac{1}{1+\xi}n_{1\,\mu}
\right)\nonumber \\
&& \times \left[\frac{1}{z_1(x+\xi-i\epsilon)} + \frac{1}{\bar
z_1(x-\xi+i\epsilon)}    \right]\,,
\end{eqnarray}
and obviously coincides with the hard part of the $\rho-$meson
electroproduction amplitude. The tensorial structure
 $2\xi\,n^\mu_{2\,} +\frac{1}{1+\xi}n^\mu_{1}=p_1^\mu + Q_1^2/s\,p_1^\mu$
coincides again with the one
present in Eq.(\ref{T}).

Passing to the TDA, let us consider the definition of
$\gamma^*_L(q_2) \to \rho^q_L(k_2) $ TDA, $T(x,\xi,t_{min})$, in
which we assume that the meson is built from a quark with a single
flavour, $\rho^q_L(k_2)= \bar q q$. The vector $P=1/2(q_2
+k_2)\approx n_2$ in our kinematics, and the ray-vector of
coordinates is oriented along the light-cone vector
$n=n_1/(n_1.n_2)$. The non-local correlator defining the TDA is
given by the formula
\begin{eqnarray}
\label{defTDA}
&&\int\,\frac{dz^-}{2\pi}\,e^{ix(P.z)}\,\langle \rho^q_L(k_2)|
\bar q(-z/2)\hat n\,e^{-ieQ_q\,\int\limits_{z/2}^{-z/2}\,
dy_\mu\,A^\mu(y) }q(z/2)|\gamma^*(q_2)\rangle
\nonumber \\
&&= \frac{e\,Q_q\,f_\rho}{P^+}\,\frac{2}{Q_2^2}\,
\epsilon_\nu(q_2)\left((1+\xi)n_2^\nu +\frac{Q_2^2}{s(1+\xi)}n_1^\nu\right)\,T(x,\xi,t_{min})\;,
\end{eqnarray}
in which we explicitly show the electromagnetic Wilson-line assuring
the abelian gauge invariance of the non-local operator. On the
contrary, for simplicity of notation, we omit the Wilson line
required by the non-abelian QCD invariance since it does not play
any role in this case.
 Note also that the factor $(1+\xi)n_2^\nu
+\frac{Q_2^2}{s(1+\xi)}n_1^\nu= p_2^\nu + \frac{Q_2^2}{s}p_1^\nu$
corresponds to a part of the tensorial structure of the second term
in Eq.~(\ref{T}).

\begin{figure}[htb]
\psfrag{p}[cc][cc]{$\slashchar{p}_2$}
\psfrag{p2}[cc][cc]{$\slashchar{p}_2$}
\psfrag{q1}[cc][cc]{$q_1$}
\psfrag{q2}[cc][cc]{$q_2$}
\psfrag{Tda}[cc][cc]{$TDA_H$}
\psfrag{n}[cc][cc]{$\slashchar{n}$}
\centerline{\scalebox{1.}
{
$\begin{array}{ccccccc}
\raisebox{-0.46 \totalheight}{\epsfig{file=TDA.eps,width=\scalingb}}
&=&
\raisebox{-0.46 \totalheight}{\epsfig{file=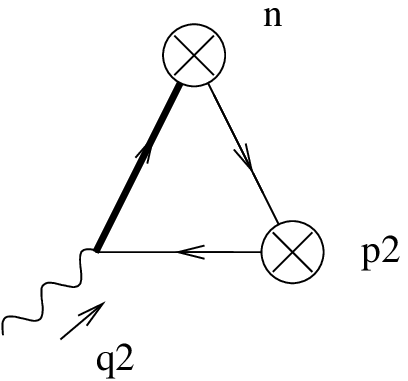,width=\scalingb}}
&+&
\raisebox{-0.46\totalheight}{\epsfig{file=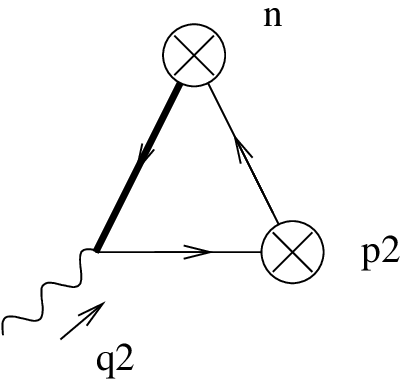,width=\scalingb}}
&+&
\raisebox{-0.42\totalheight}{\epsfig{file=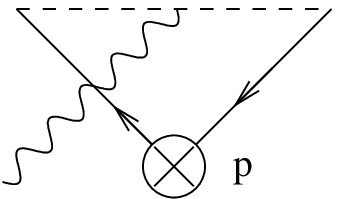,width=\scalingml}}
\end{array}
$}}
\caption{The hard part of the TDA  at  order $e \,Q_q$.
\label{devTDA}}
\end{figure}

Now, the simple perturbative calculation of the matrix element in
(\ref{defTDA}) in the lowest order in the electromagnetic coupling
constant $e$ leads to the expression for $T(x,\xi,t_{min})$ given by
Eq.~(\ref{tda}). The contributing diagrams are drawn  in
Fig.\ref{devTDA}.
In particular, the contribution to the rhs of (\ref{defTDA})
proportional to the vector $n_1^\nu$ (or $p_1^\nu$) corresponds to
the contribution coming from the expansion of the electromagnetic
Wilson line, illustrated by the last diagram in Fig.\ref{devTDA}.

Putting all factors together and restoring  the flavour structure of the $\rho^0$, we obtain the
 factorized form involving a TDA,  of the
 expression
$T^{\alpha\,\beta}p_{2\,\alpha}p_{1\,\beta}$ in Eq.~(\ref{Tlong}) as
\begin{equation}
\label{tdafactor}
T^{\alpha\,\beta}p_{2\,\alpha}p_{1\,\beta}=
-if_\rho^2e^2(Q_u^2+Q_d^2)g^2\,\frac{C_F}{8N_c}\int\limits_{-1}^1 dx\,\int\limits_0^1 dz_1\,
\left[\frac{1}{\bar z_1(x-\xi)} + \frac{1}{z_1(x+\xi)} \right]\,T(x,\xi,t_{min})\;.
\end{equation}

Note that in this perturbative analysis, only the DGLAP part of the
TDA, with \\$1~\ge~|x|~\ge~\xi$, contributes. This is a consequence
of the support properties of the  $\rho-$meson distribution
amplitude.

\section{Conclusions}

Thus  the perturbative analysis of the process
$\gamma^*\;\gamma^*\to \rho^0_L \rho^0_L$ in the Born approximation
leads to two different types of QCD factorization. We have shown
that
 the
polarization states of the photons dictate either the factorization involving a GDA or involving a
TDA. Usually these two types of factorizations are applied to two different kinematical regimes.
The arbitrariness in choosing values of photon virtualities $Q^2_i$ shows that there may exist
an intersection region where both types of factorization are simultaneously valid.

We have restricted our analysis to the case of longitudinally polarized $\rho^0-$mesons. In this way
we have avoided the potential problems
\cite{QCDbreaking}
due to the breaking of QCD factorization with GDA or TDA at the end-point region
of the distribution amplitudes of transversally polarized vector mesons.
It would be interesting to find out, whether a
factorizing formula with GDA or TDA can also be obtained in this case.

Although we have restricted ourselves, for simplicity, to the forward kinematics, we believe that
similar results may be obtained in a more general case.
Also, we would like to mention that our results involving $\rho^0-$mesons can be generalized to the case of the
production of
a $(\rho^+ \rho^-)-$meson pair. On the theoretical side, in order to
preserve the electromagnetic gauge invariance of the TDA $\gamma^* \to \rho^\pm$,
one should modify the definition of the non-local
correlator (\ref{defTDA}). This may be done by applying the Mandelstam approach \cite{Mand}, i.e. by  replacing
the electromagnetically gauge invariant corelator (\ref{defTDA}) by the product of two effective electromagnetically gauge
invariant quark fields
 $q(z)\exp (i\,e_q\int\limits_z^\infty \,dy\,n_\mu A^\mu(y))$.

\section*{Acknowledgments}

This  work  is partially supported by the Polish Grant 1 P03B 028
28, the French-Polish scientific agreement Polonium, the triangular
program ECO-NET, and the european  Joint Research Activity I3
Program Hadronic Physics RII3-CT-506078. L.Sz.\ is a Visiting Fellow
of the Fonds National pour la Recherche Scientifique (Belgium).
We thank C.~Ewerz, G.~Korchemsky, J-P.~Lansberg, D.~M\"uller and A.~Shuvaev for discussions.



\begin{thebibliography}{99}

\bibitem{PSW}
  B.~Pire, L.~Szymanowski and S.~Wallon,
  Eur.\ Phys.\ J.\ C {\bf 44} (2005) 545
  [arXiv:hep-ph/0507038].


\bibitem{EPSW}
  R.~Enberg, B.~Pire, L.~Szymanowski and S.~Wallon,
  Eur.\ Phys.\ J.\ C {\bf 45} (2006) 759
  [arXiv:hep-ph/0508134].






\bibitem{BLphysrev24}
  S.~J.~Brodsky and G.~P.~Lepage,
  Phys.\ Rev.\ D {\bf 24} (1981) 1808.

\bibitem{ERBL}
G.P. Lepage and S.J. Brodsky, Phys.\ Lett.\ {\bf B87},
359 (1979); A.V. Efremov and A.V. Radyushkin, Phys.\ Lett.\ {\bf B94},
245 (1980).


\bibitem{DFKV}
  M.~Diehl, T.~Feldmann, P.~Kroll and C.~Vogt,
  Phys.\ Rev.\ D {\bf 61} (2000) 074029
  [arXiv:hep-ph/9912364].

 \bibitem{DM}
D.~M{\"u}ller, D.~Robaschik, B.~Geyer, F.~M.~Dittes, J.~Ho\v{r}ej\v{s}i,
Fortsch.\ Phys.\  {\bf 42},  101 (1994);



 \bibitem{GDA}
M.~Diehl, T.~Gousset, B.~Pire and O.~Teryaev,
Phys.\ Rev.\ Lett.\  {\bf 81}, 1782 (1998);
M.~Diehl, T.~Gousset and B.~Pire,
Phys.\ Rev.\ D {\bf 62}, 073014 (2000);
I.~V.~Anikin, B.~Pire and O.~V.~Teryaev,
Phys.\ Rev.\ D {\bf 69}, 014018 (2004).





\bibitem{TDA}
  B.~Pire and L.~Szymanowski,
  Phys.\ Rev.\ D {\bf 71} (2005) 111501;
  Phys.\ Lett.\ B {\bf 622} (2005) 83;
  J.~P.~Lansberg, B.~Pire and L.~Szymanowski,
  Phys.\ Rev.\ D {\bf 73} (2006) 074014
  [arXiv:hep-ph/0602195].



\bibitem{GPD}
X.~Ji,
Phys.\ Rev.\ Lett.\  {\bf 78}, 610 (1997);  Phys.\
  Rev.\ {\bf D55}, 7114 (1997);
A.~V.~Radyushkin,
Phys.\ Rev.\ D {\bf 56}, 5524 (1997);
J. Bl\"umlein, B. Geyer and D. Robaschik,
Phys.\ Lett.\ {\bf B406}, 161 (1997);
For a review, see M.~Diehl,
Phys.\ Rept.\  {\bf 388}, 41 (2003)
 and references therein.


\bibitem{QCDbreaking}
  L.~Mankiewicz and G.~Piller,
  Phys.\ Rev.\ D {\bf 61} (2000) 074013
  [arXiv:hep-ph/9905287];
  I.~V.~Anikin and O.~V.~Teryaev,
  Phys.\ Lett.\ B {\bf 554} (2003) 51
  [arXiv:hep-ph/0211028].

\bibitem{Mand}
  S.~Mandelstam,
  Annals Phys.\  {\bf 19}, 1 (1962).

\end{thebibliography}
\end{document}